\documentclass{elsart}
\usepackage{epsfig}

\begin{document}

\begin{frontmatter}

\title{Comparison between the probability distribution of returns in
the Heston model and empirical data for stock indices}

\author{A. Christian Silva, Victor M. Yakovenko\thanksref{email}}

\thanks[email]{yakovenk@physics.umd.edu,
   http://www2.physics.umd.edu/\~{}yakovenk}

\address{Department of Physics, University of Maryland, College Park,
  MD 20742-4111, {\bf cond-mat/0211050, 4 November 2002} \hfill
  USA}


\begin{abstract}
  We compare the probability distribution of returns for the three
  major stock-market indices (Nasdaq, S\&P500, and Dow-Jones) with an
  analytical formula recently derived by Dr\u{a}gulescu and Yakovenko
  for the Heston model with stochastic variance.  For the period of
  1982--1999, we find a very good agreement between the theory and the
  data for a wide range of time lags from 1 to 250 days.  On the other
  hand, deviations start to appear when the data for 2000--2002 are
  included.  We interpret this as a statistical evidence of the major
  change in the market from a positive growth rate in 1980s and 1990s
  to a negative rate in 2000s.
\end{abstract}

\begin{keyword}
Econophysics \sep Stochastic Volatility \sep Heston \sep Stock Market
Returns \PACS  02.50.-r \sep 89.65.-s
\end{keyword}

\end{frontmatter}

\section{Introduction}

Models of multiplicative Brownian motion with stochastic volatility
have been a subject of extensive studies in finance, particularly in
relation with option pricing \cite{Papanicolaou}.  One of the popular
models is the so-called Heston model \cite{Heston}, for which many
exact mathematical results can be obtained.  Recently, Dr\u{a}gulescu
and Yakovenko (DY) \cite{Yakovenko} derived a closed analytical
formula for the probability distribution function (PDF) of log-returns
in the Heston model.  They found an excellent agreement between the
formula and the empirical data for the Dow-Jones index for the period
of 1982--2001.  (Discussion of other work on returns distribution and
references can be found in Ref.\ \cite{Yakovenko}.)

In the present paper, we extend the comparison by including the data
for Nasdaq and S\&P500.  We find that the DY formula agrees very well
with the data for the period of 1982--1999.  However, when the data
for 2000-2002 are included, systematic deviations are observed, which
reflect a switch of the market from upward to downward trend around
2000.

\section{Probability distribution of log-returns in the Heston model}

In this Section, we briefly summarize the results of the DY paper
\cite{Yakovenko}.  Let us consider a stock, whose price $S_t$, as a
function of time $t$, obeys the stochastic differential equation of
multiplicative Brownian motion:
\begin{equation}\label{eqS}
  dS_t = \mu S_t\, dt + \sigma_t S_t\, dW_t^{(1)}.
\end{equation}
Here the subscript $t$ indicates time dependence, $\mu$ is the drift
parameter, $W_t^{(1)}$ is a standard random Wiener process, and
$\sigma_t$ is the time-dependent volatility.  Changing the variable in
(\ref{eqS}) from price $S_t$ to log-return $r_t=\ln(S_t/S_0)$ and
eliminating the drift by introducing $x_t=r_t-\mu t$, we find:
\begin{equation}\label{eqX}
  dx_t = - \frac{v_t}{2}\,dt + \sqrt{v_t}\,dW_t^{(1)}.
\end{equation}
where $v_t=\sigma_t^2$ is the variance.

Let us assume that the variance $v_t$ obeys the following
mean-reverting stochastic differential equation:
\begin{equation} \label{eqVar}
  dv_t = -\gamma(v_t - \theta)\,dt + \kappa\sqrt{v_t}\,dW_t^{(2)}.
\end{equation}
Here $\theta$ is the long-time mean of $v$, $\gamma$ is the rate of
relaxation to this mean, $W_t^{(2)}$ is a standard Wiener process, and
$\kappa$ is the variance noise.  In general, the Wiener process in
(\ref{eqVar}) may be correlated with the Wiener process in
(\ref{eqS}):
\begin{equation} \label{rho}
  dW_t^{(2)} = \rho\,dW_t^{(1)} + \sqrt{1-\rho^2}\,dZ_t,
\end{equation}
where $Z_t$ is a Wiener process independent of $W_t^{(1)}$, and
$\rho\in[-1,1]$ is the correlation coefficient.

The coupled stochastic processes (\ref{eqX}) and (\ref{eqVar})
constitute the Heston model \cite{Heston}.  In a standard manner
\cite{Gardiner}, the Fokker-Planck equation can be derived for the
transition probability $P_t(x,v\,|\,v_i)$ to have log-return $x$ and
variance $v$ at time $t$ given the initial log-return $x=0$ and
variance $v_i$ at $t=0$:
\begin{eqnarray}
  \frac{\partial}{\partial t}P &=& 
     \gamma\frac{\partial}{\partial v}\left[(v-\theta)P\right]
     + \frac12\frac{\partial}{\partial x}(vP)
\label{FP} \\ 
  && {} +\rho\kappa\frac{\partial^2}{\partial x\,\partial v}(vP)
     +\frac12\frac{\partial^2}{\partial x^2}(vP)  
     +\frac{\kappa^2}{2}\frac{\partial^2}{\partial v^2}(vP).  
\nonumber
\end{eqnarray}
A general analytical solution of Eq.\ (\ref{FP}) for
$P_t(x,v\,|\,v_i)$ was obtained in Ref.\ \cite{Yakovenko}.  Then
$P_t(x,v\,|\,v_i)$ was integrated over the final variance $v$ and
averaged over the stationary distribution $\Pi_\ast(v_i)$ of the
initial variance $v_i$:
\begin{equation} \label{P_t(x)}
  P_t(x)= \int_0^\infty dv_i\int_0^\infty dv\,
  P_t(x,v\,|\,v_i)\,\Pi_\ast(v_i).
\end{equation}
The function $P_t(x)$ in Eq.\ (\ref{P_t(x)}) is the PDF of log-returns
$x$ after the time lag $t$.  It can be direct compared with financial
data.  It was found in Ref.\ \cite{Yakovenko} that data fits are not
very sensitive to the parameter $\rho$, so below we consider only the
case $\rho=0$ for simplicity.

The final expression for $P_t(x)$ at $\rho=0$ (the DY formula
\cite{Yakovenko}) has the form of a Fourier integral:
\begin{eqnarray} 
  && P_t(x) = \frac{e^{-x/2}}{x_0}\int_{-\infty}^{+\infty}
  \frac{d\tilde p}{2\pi} \,
  e^{i\tilde p\tilde x + F_{\tilde t}(\tilde p)},
\label{Pfinal'-} \\
  && F_{\tilde t}(\tilde p)=\frac{\alpha\tilde t}{2}
  - \alpha\ln\left[\cosh\frac{\tilde\Omega\tilde t}{2} +
  \frac{\tilde\Omega^2+1}{2\tilde\Omega}
  \sinh\frac{\tilde\Omega\tilde t}{2}\right],
\label{phaseF'-} \\
  && \tilde\Omega=\sqrt{1+\tilde p^2}, \quad \tilde t=\gamma t, \quad 
  \tilde x=x/x_0, \quad x_0=\kappa/\gamma, \quad 
  \alpha=2\gamma\theta/\kappa^2.
\end{eqnarray}
In the long-time limit $\tilde t\gg2$, Eqs.\ (\ref{Pfinal'-}) and
(\ref{phaseF'-}) exhibit scaling behavior, i.e.\ $P_t(x)$ becomes a
function of a single combination $z$ of the two variable $x$ and $t$
(up to the trivial normalization factor $N_t$ and unimportant factor
$e^{-x/2}$):
\begin{eqnarray} 
  && P_t(x)=N_t\,e^{-x/2}P_{\ast}(z),\quad P_{\ast}(z)=K_1(z)/z, \quad
  z=\sqrt{\tilde x^2 + \bar t^2},
\label{Pbess-} \\
  && \bar t=\alpha\tilde t/2=t\theta/x_0^2, \quad
  N_t=\bar te^{\bar t}/\pi x_0,
\end{eqnarray}
where $K_1(z)$ is the first-order modified Bessel function.

\section{Comparison between the DY theory and the data}

\begin{figure}
\centerline{
\epsfig{file=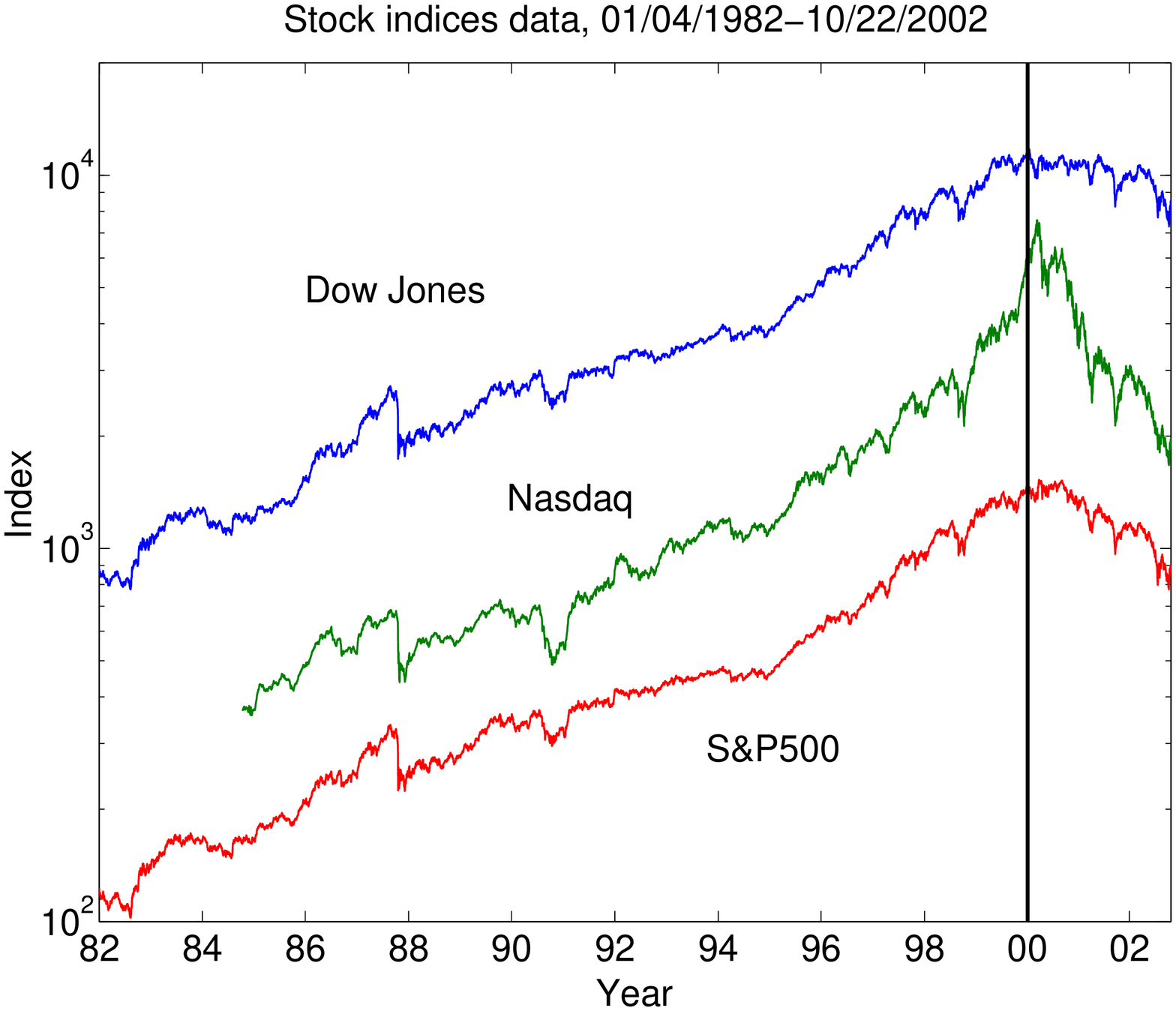,width=0.43\linewidth,angle=-90}
\epsfig{file=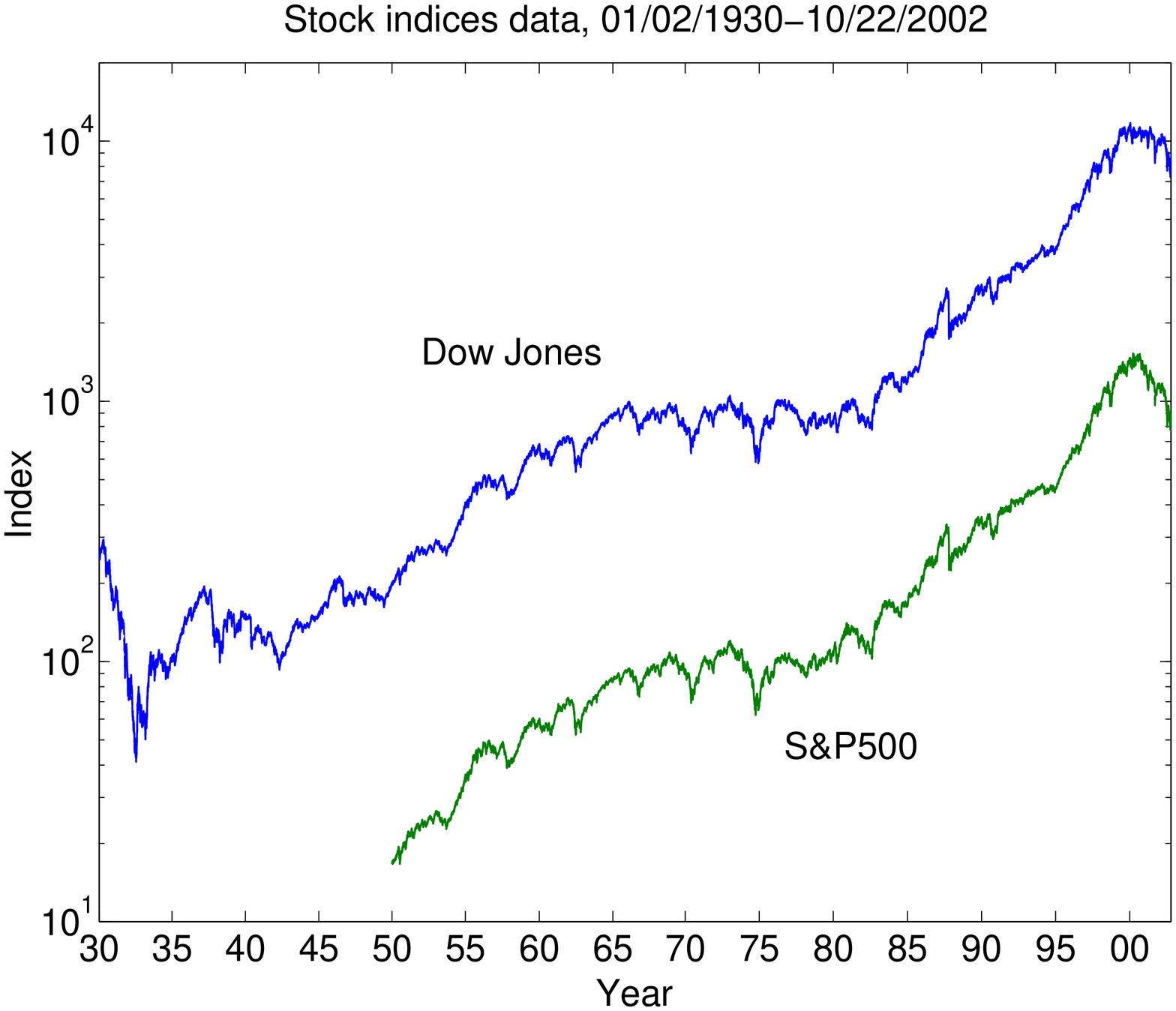,width=0.43\linewidth,angle=-90}}
\caption{\footnotesize\sf Historical evolution of the three major
  stock-market indices, shown in the log-linear scale.  The Nasdaq
  curve is shifted up by the factor of 1.5 for clarity.  The vertical
  line separates the regions with the average positive and negative
  growth rates.}
\label{fig:timeseries}
\end{figure}

We analyzed the data for the three major stock-market indices:
Dow-Jones, S\&P500, and Nasdaq.  From the Yahoo Web site \cite{Yahoo},
we downloaded the daily closing values of Dow-Jones and S\&P500 from 4
January 1982 to 22 October 2002 and all available data for Nasdaq from
11 October 1984 to 22 October 2002.  The downloaded time series
$\{S_\tau\}$ are shown in the left panel of Fig.\
\ref{fig:timeseries}.  It is clear that during 1980s and 1990s all
three indices had positive exponential growth rates, followed by
negative rates in 2000s.  For comparison, in the right panel of Fig.\
\ref{fig:timeseries}, we show the time series from 1930 to 2002.
Contrary to the mutual-funds propaganda, stock market does not always
increase.  During 1930s (Great Depression) and 1960s--1970s
(Stagnation), the average growth rate was zero or negative.  One may
notice that such fundamental changes of the market trend occur on a
very long time scale of the order of 15--20 years.

\begin{table}[b]
\caption{\footnotesize\sf Parameters of the Heston
  model obtained from the fits of the Nasdaq, S\&P500, and Dow-Jones
  data from 1982 to 1999 using $\rho=0$ for the correlation
  coefficient. \label{paramVal}}
\begin{tabular}{c|cccccccc}
\hline
& $\gamma$ & $1/\gamma$ & $\theta$ & $\kappa$ & $\mu$ & $\bar\mu$ 
  & $\alpha$ & $x_{0}$ \\
& ${1\over{\rm year}}$ & day & ${1\over{\rm year}}$ & 
${1\over{\rm year}}$ & ${1\over{\rm year}}$ & ${1\over{\rm year}}$ & & \\
\hline
Nasdaq & 114 & 2.2 & 3.6\% & 5.3 & 16\% & 14\% & 0.3 & 4.7\% \\
\hline
S\&P500 & 17 & 15 & 1.8\% & 0.67 & 13\% & 12\% & 1.36 & 4.0\% \\
\hline
Dow-Jones & 24 & 10 & 2\% & 0.94 & 14\% & 13\% & 1.1 & 3.9\% \\
\hline
\end{tabular}
\end{table}

Using the procedure described in Ref.\ \cite{Yakovenko}, we extract
the PDFs $P_t^{\rm(data)}(r)$ of log-returns $r$ for different time
lags $t$ from the time series $\{S_\tau\}$ for all three indices.  In
the DY theory \cite{Yakovenko}, the actual (empirically observed)
growth rate $\bar\mu$ is related to the bare parameter $\mu$ by the
following relation: $\bar\mu=\mu-\theta/2$, and $P_t^{\rm(data)}(x)$
is obtained by replacing the argument $r\to x+\mu t$.  The parameters
$\bar\mu$ were found by fitting the time series in the left panel of
Fig.\ \ref{fig:timeseries} to straight lines.  With the constraint
$\mu=\bar\mu+\theta/2$, the other parameters of the Heston model
($\gamma$, $\theta$, $\kappa$) were obtained by minimizing the
mean-square deviation $\sum_{x,t}|\ln P_t^{\rm(data)}(x)-\ln
P_t(x)|^2$ between the empirical data and the DY formula
(\ref{Pfinal'-}) and (\ref{phaseF'-}), with the sum taken over all
available $x$ and the time lags $t=1$, 5, 20, 40, and 250 days.  This
procedure was applied to the data from 1982 (1984 for Nasdaq) to 31
December 1999, and the values of the obtained parameters are shown in
Table \ref{paramVal}.  The model parameters for Dow-Jones and S\&P500
are similar, whereas some parameters for Nasdaq are significantly
different.  Namely, the variance relaxation time $1/\gamma$ is much
shorter, the variance noise $\kappa$ is much bigger, and the parameter
$\alpha$ is much smaller for Nasdaq.  All of this is consistent with
the general notion that Nasdaq is more volatile than Dow-Jones and
S\&P500. On the other hand, the average growth rates $\bar\mu$ of all
three indices are about the same, so the greater risk in Nasdaq does
not result in a higher average return.

\begin{figure}
\centerline{
  \epsfig{file=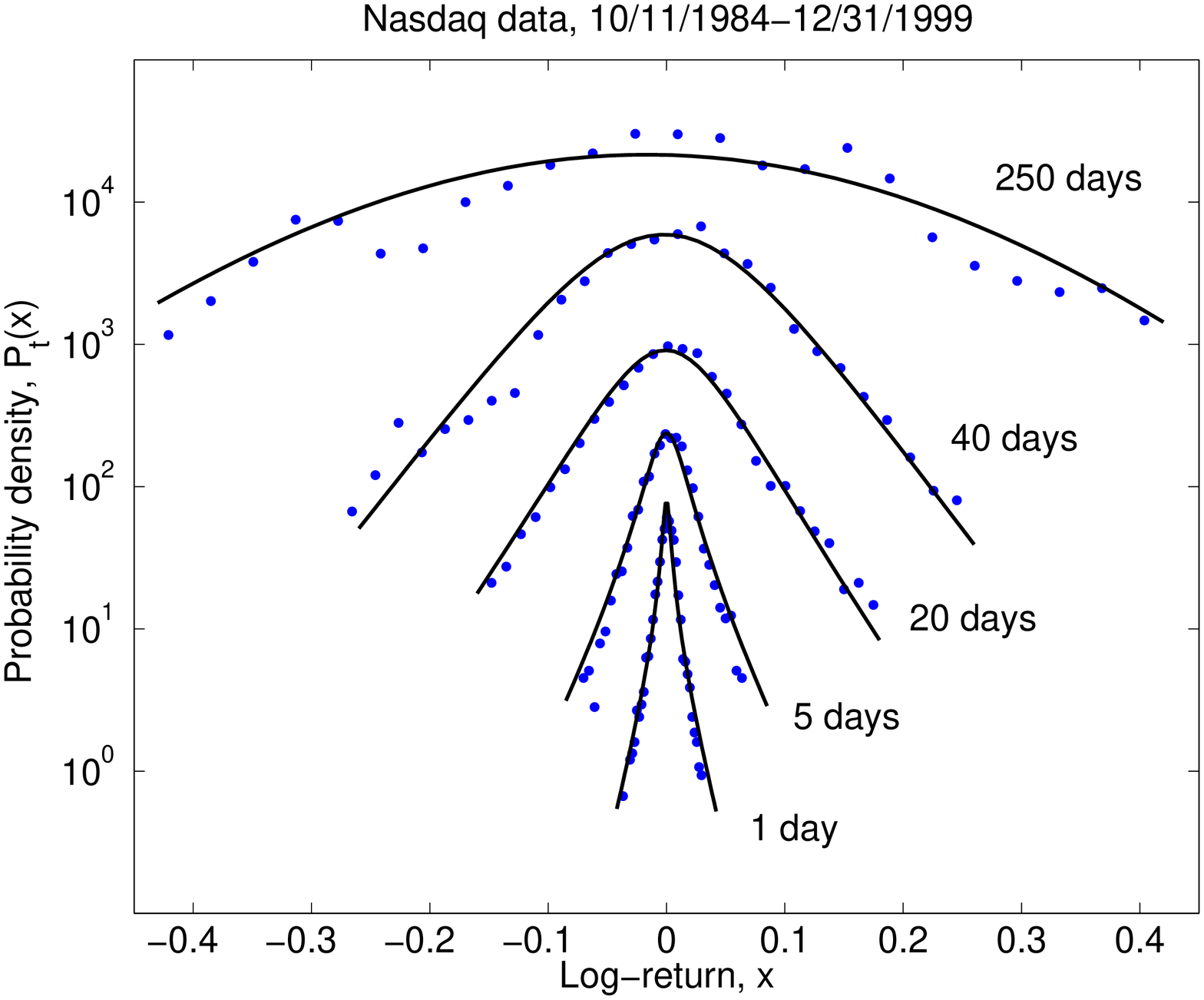,width=0.41\linewidth,angle=-90}
  \epsfig{file=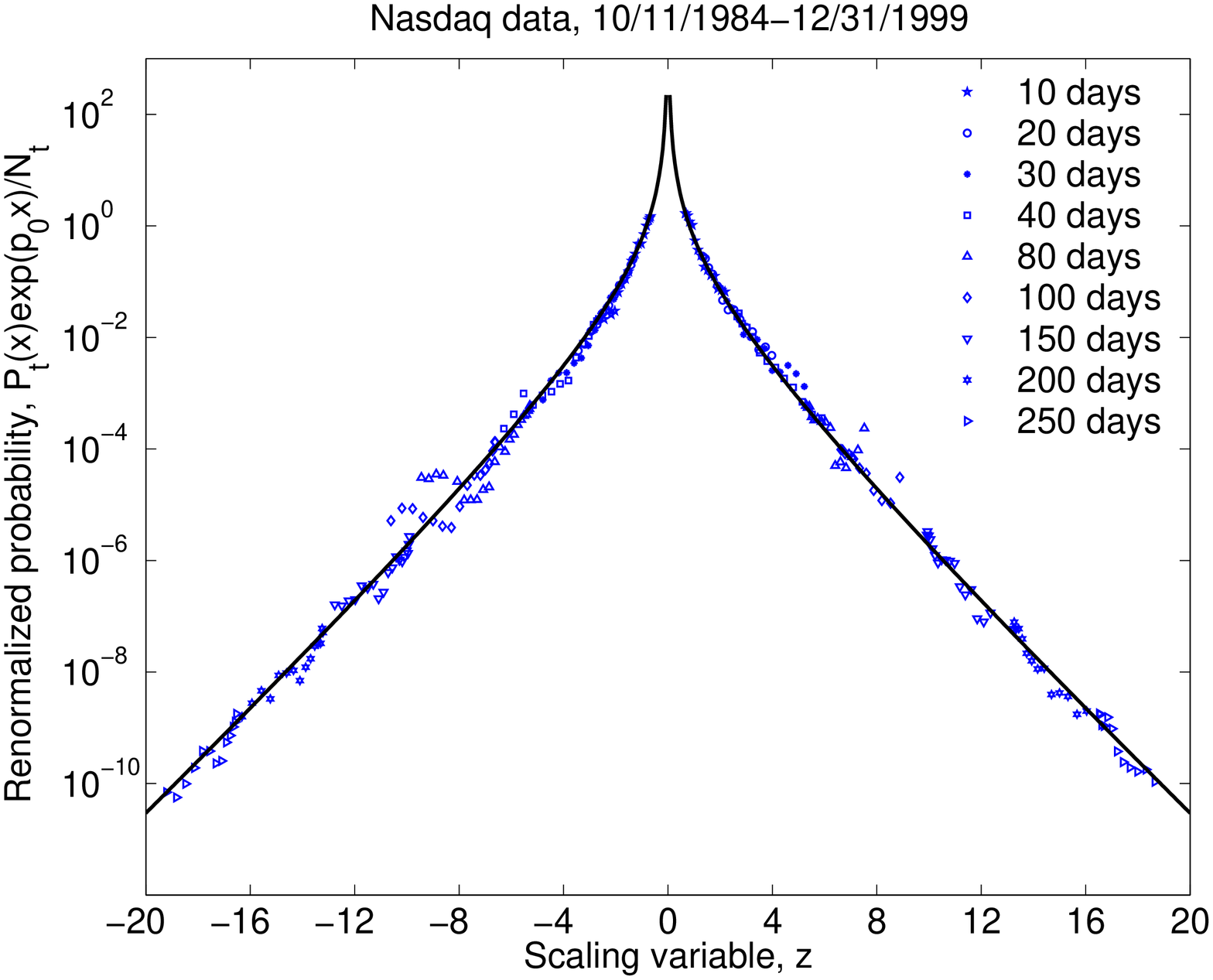,width=0.41\linewidth,angle=-90}}
\caption{\footnotesize\sf Comparison between the 1984--1999 Nasdaq
  data (points) and the Dr\u{a}gulescu--Yakovenko theory
  \cite{Yakovenko} (curves).  Left panel: PDFs $P_t(x)$ of log-returns
  $x$ for different time lags $t$ shifted up by the factor of 10 each
  for clarity.  Right panel: Renormalized PDF $P_t(x)e^{x/2}/N_t$
  plotted as a function of the scaling argument $z$ given in Eq.\
  (\ref{Pbess-}).  The solid line is the scaling function
  $P_{\ast}(z)=K_1(z)/z$ from Eq.\ (\ref{Pbess-}), where $K_1$ is the
  first-order modified Bessel function.}
\label{fig:NQ00}
\end{figure}

\begin{figure}[b]
\centerline{
  \epsfig{file=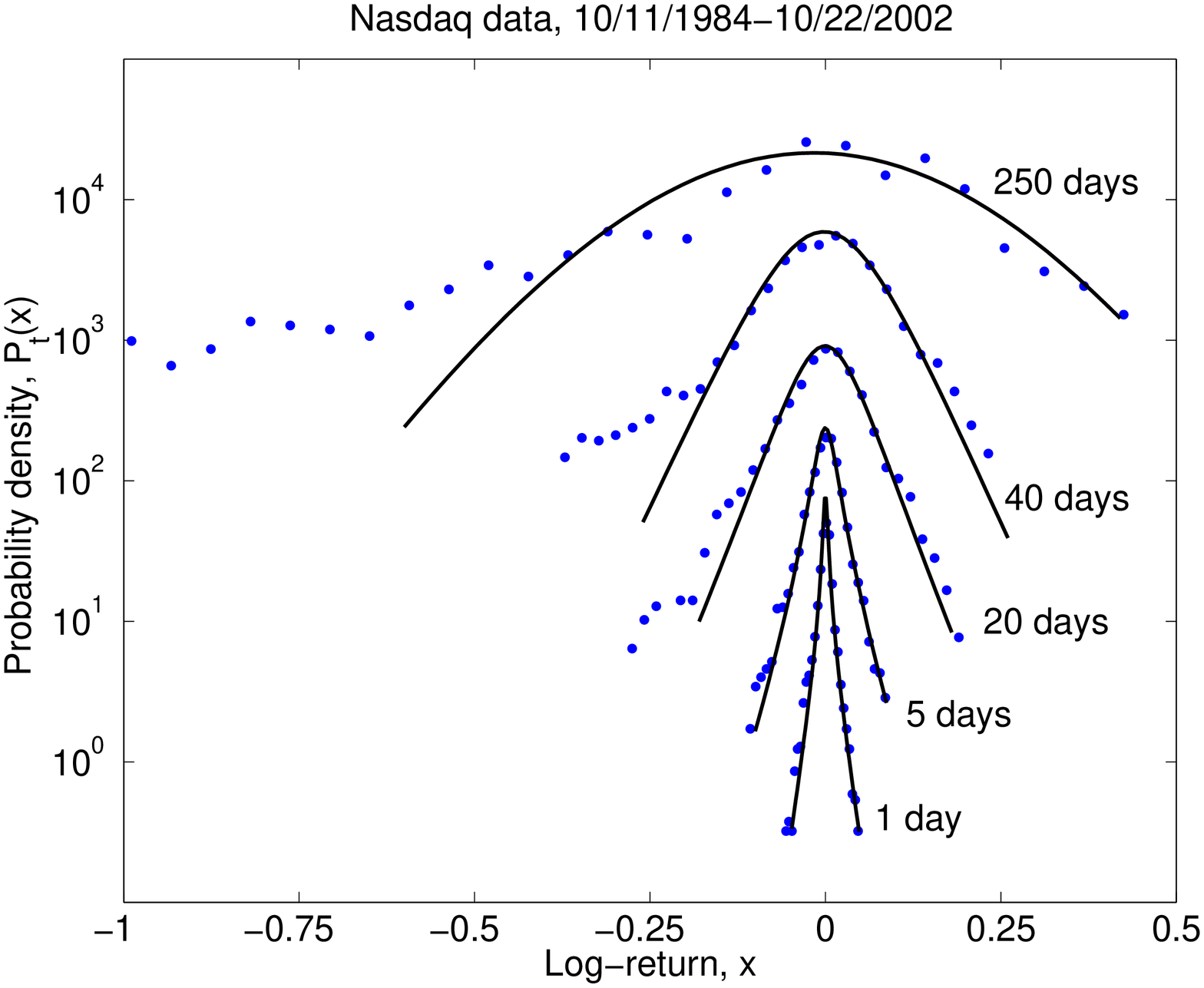,width=0.41\linewidth,angle=-90}
  \epsfig{file=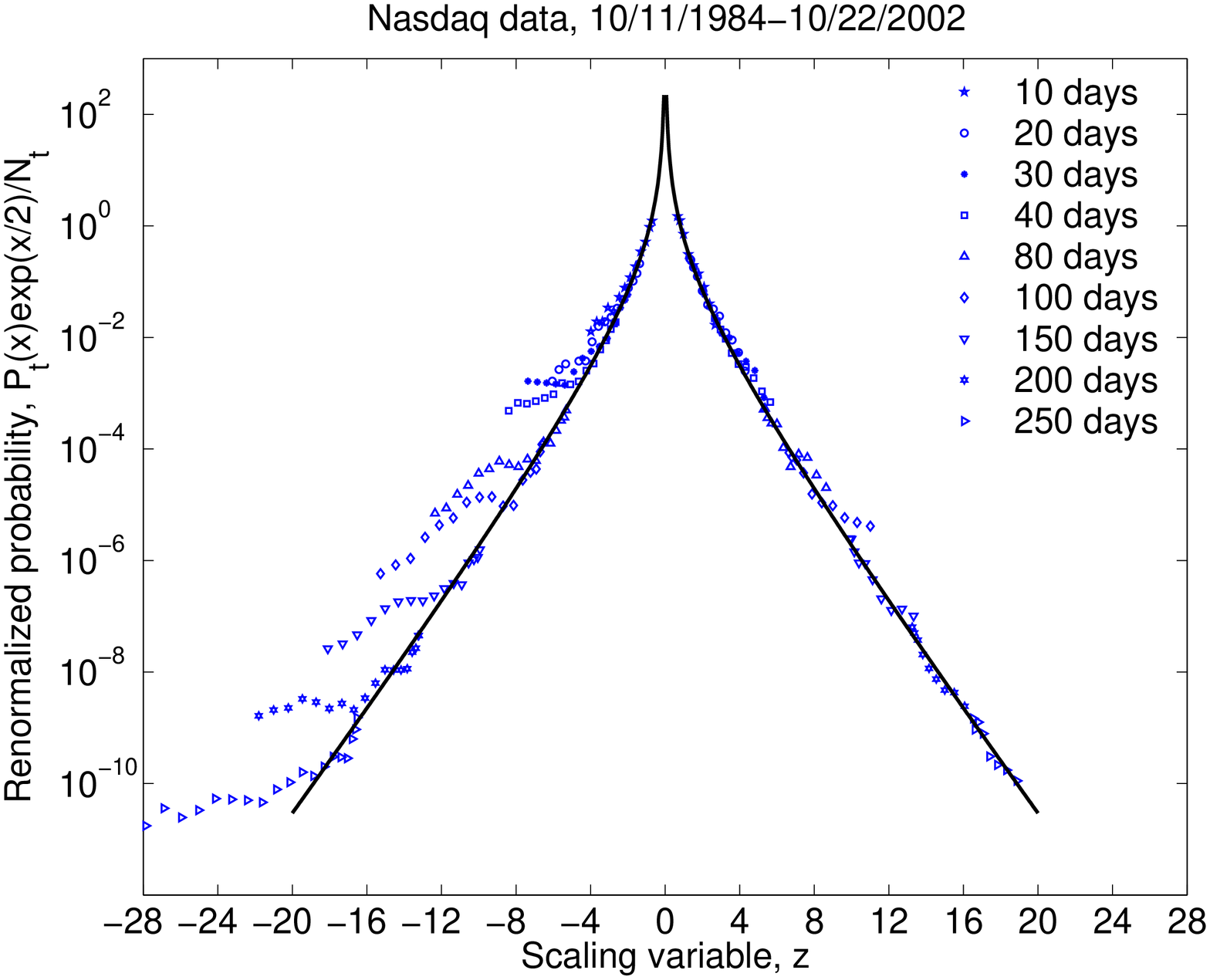,width=0.41\linewidth,angle=-90}}
\caption{\footnotesize\sf The same as in Fig.\ \ref{fig:NQ00} for
  1984--2002.}
\label{fig:NQ02}
\end{figure}

Fig.\ \ref{fig:NQ00} compares the 1984--1999 data for Nasdaq (points)
with the DY theory (curves).  The left panel shows the PDFs $P_t(x)$
(\ref{Pfinal'-}) for several time lags $t$, and the right panel
demonstrates the scaling behavior (\ref{Pbess-}).  The overall
agreement is quite good.  Particularly impressive is the scaling plot,
where the points for different time lags collapse on a single
nontrivial scaling curve spanning 10 (!) orders of magnitude.  On the
other hand, when we include the data up to 22 October 2002, the points
visibly run off the theoretical curves, as shown in Fig.\
\ref{fig:NQ02}.  We use the same values of the parameters ($\mu$,
$\gamma$, $\theta$, $\kappa$) in Fig.\ \ref{fig:NQ02} as in Fig.\
\ref{fig:NQ00}, because attempts to adjust the parameters do not
reduce the discrepancy between theory and data.  The origin of the
discrepancy is discussed in Sec.\ \ref{sec:conclusions}.

\begin{figure}
\centerline{
  \epsfig{file=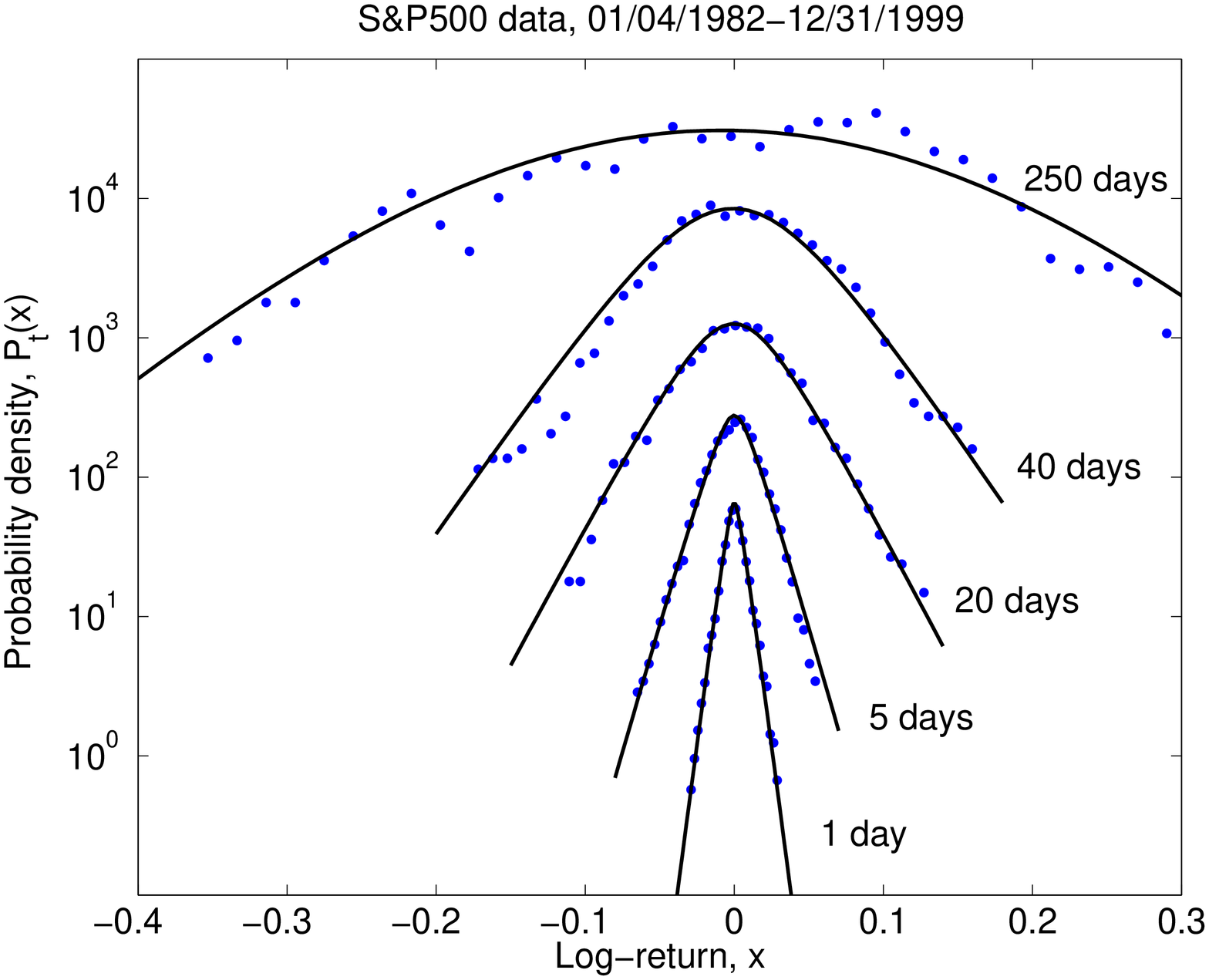,width=0.41\linewidth,angle=-90}
  \epsfig{file=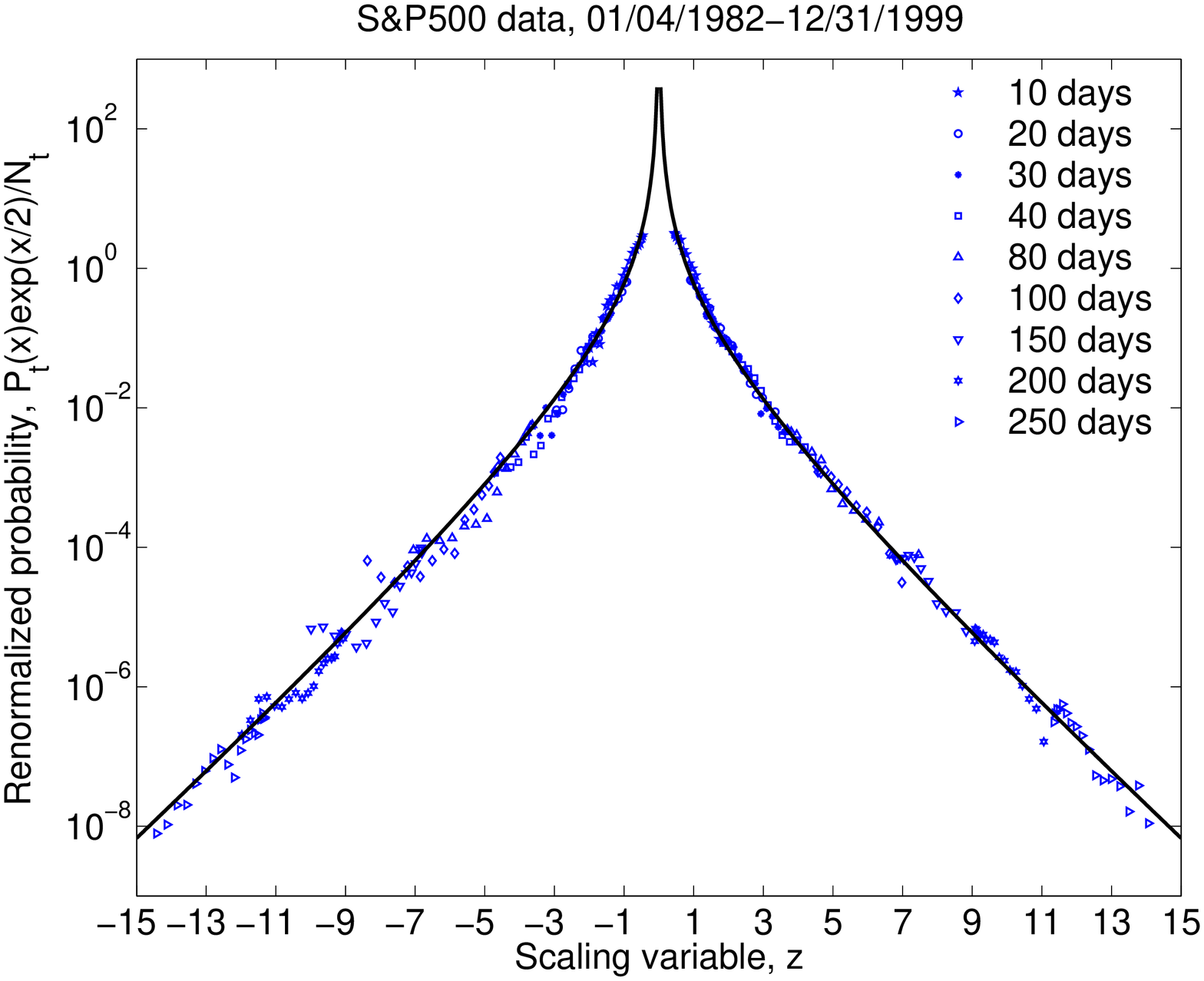,width=0.41\linewidth,angle=-90}}
\caption{\footnotesize\sf The same as in Fig.\ \ref{fig:NQ00} for
  S\&P500 for 1982--1999.}
\label{fig:SP00}
\end{figure}

\begin{figure}[b]
\centerline{
  \epsfig{file=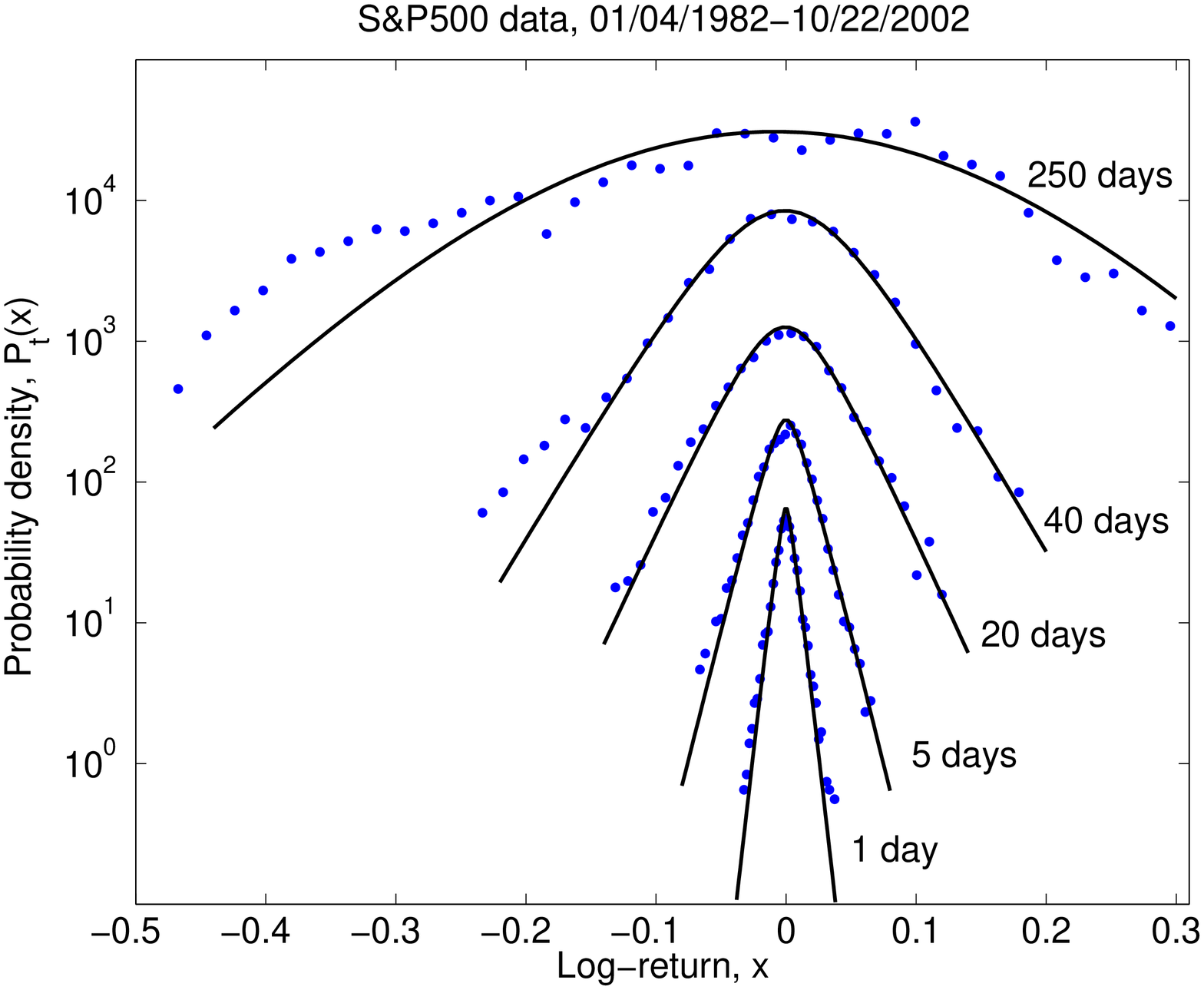,width=0.41\linewidth,angle=-90}
  \epsfig{file=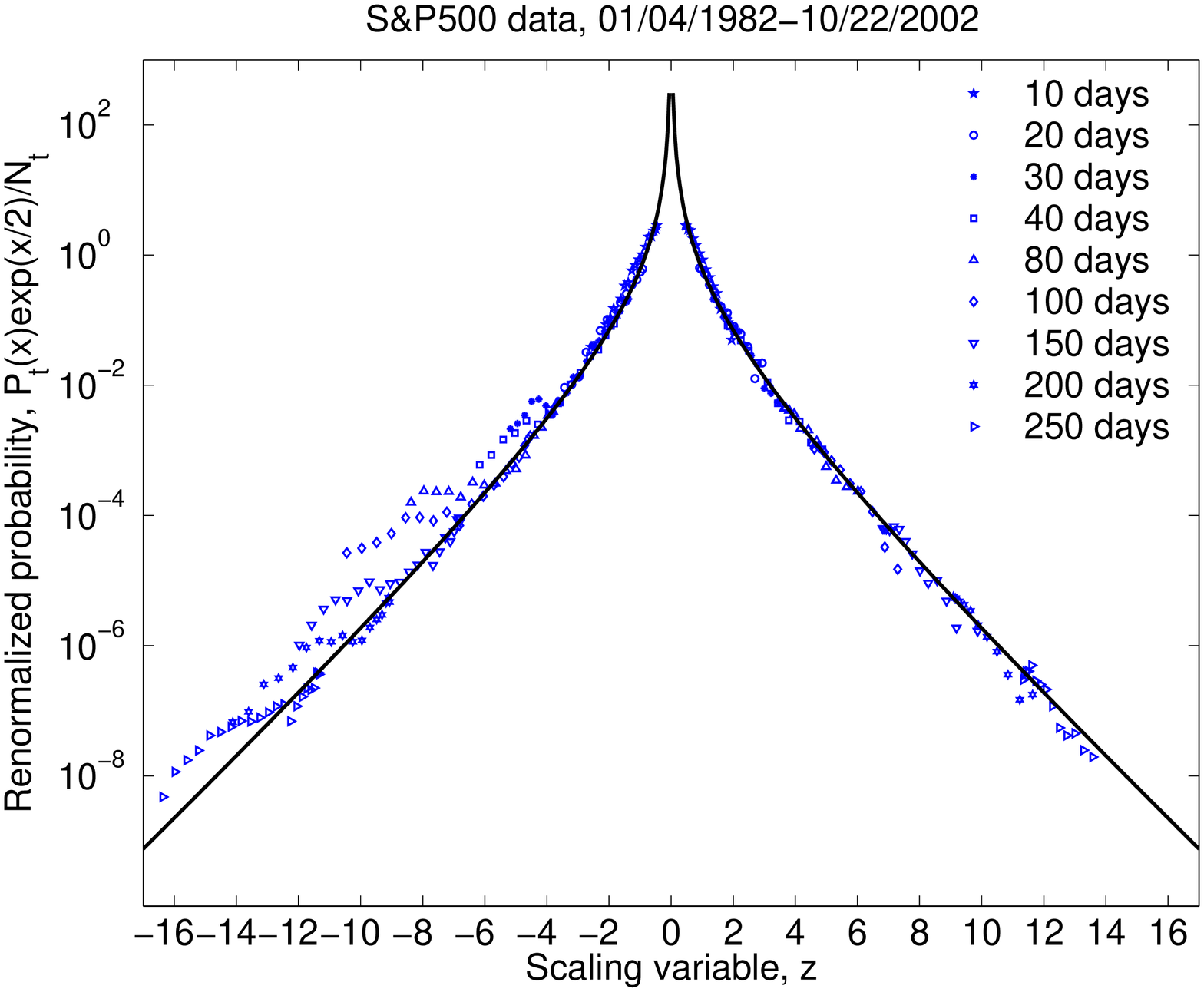,width=0.41\linewidth,angle=-90}}
\caption{\footnotesize\sf The same as in Fig.\ \ref{fig:SP00} for
  1982--2002.}
\label{fig:SP02}
\end{figure}

\begin{figure}
\centerline{
  \epsfig{file=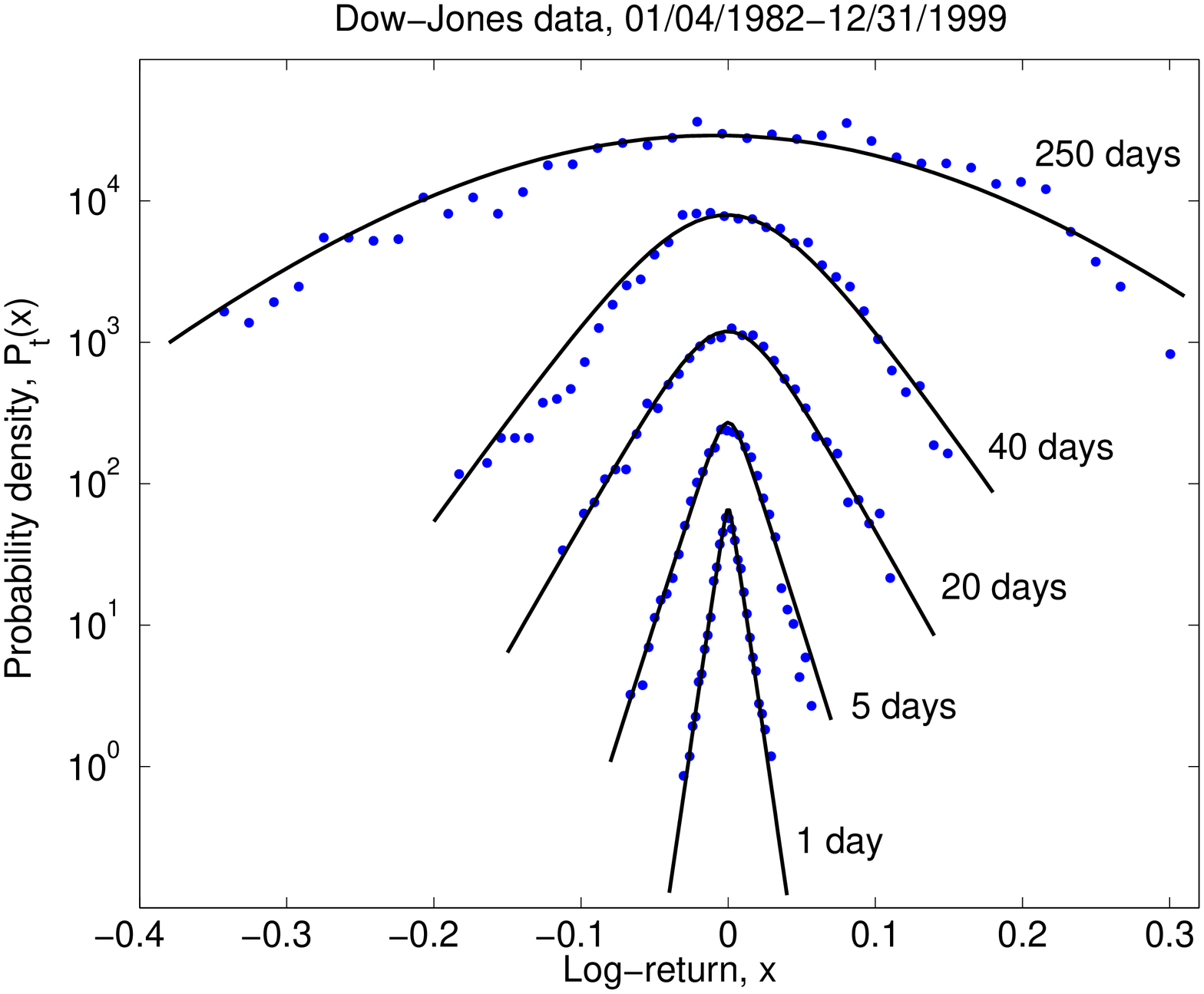,width=0.41\linewidth,angle=-90}
  \epsfig{file=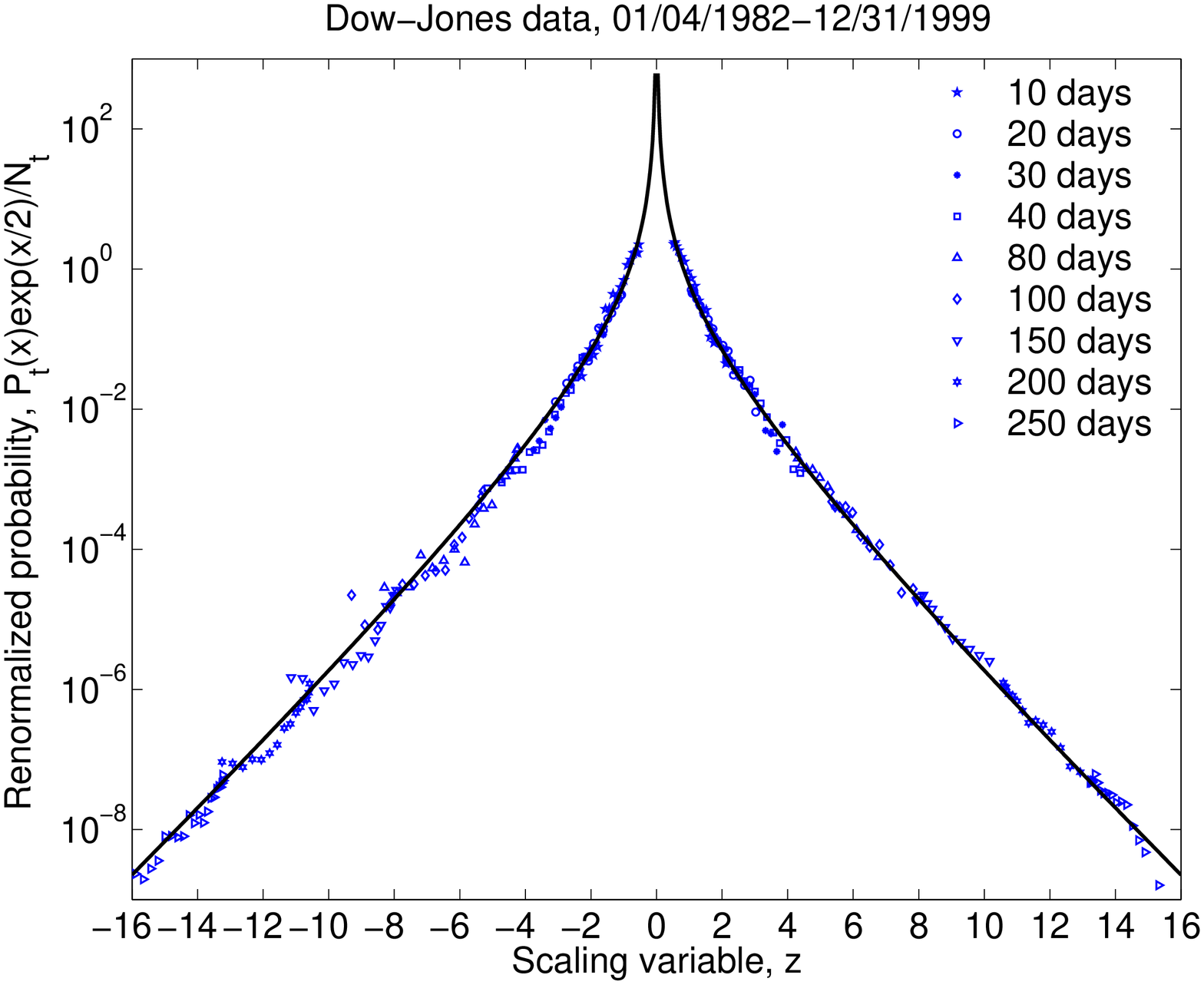,width=0.41\linewidth,angle=-90}}
\caption{\footnotesize\sf The same as in Fig.\ \ref{fig:SP00} for
  Dow-Jones.}
\label{fig:DJ00}
\end{figure}

\begin{figure}[b]
\centerline{
  \epsfig{file=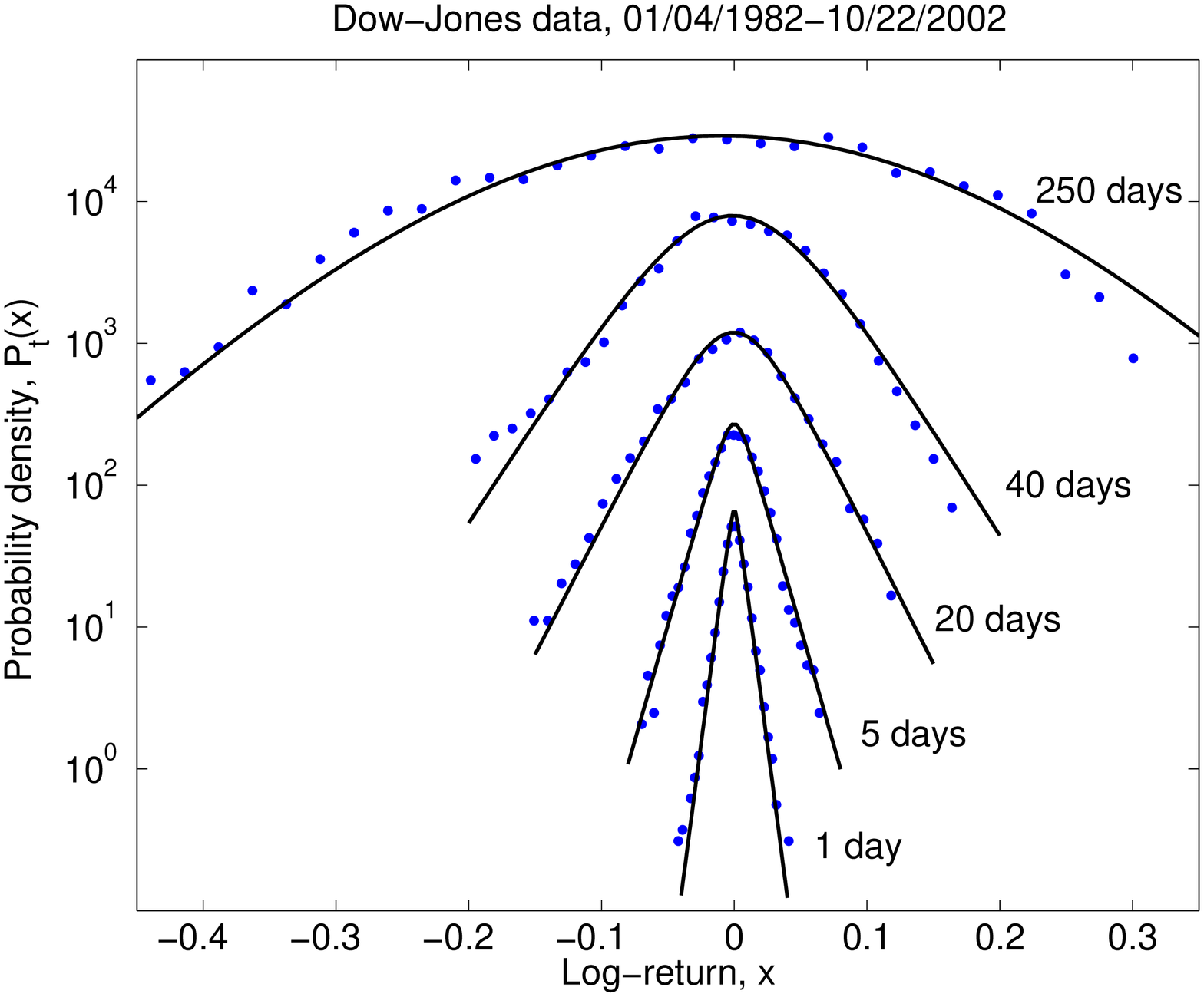,width=0.41\linewidth,angle=-90}
  \epsfig{file=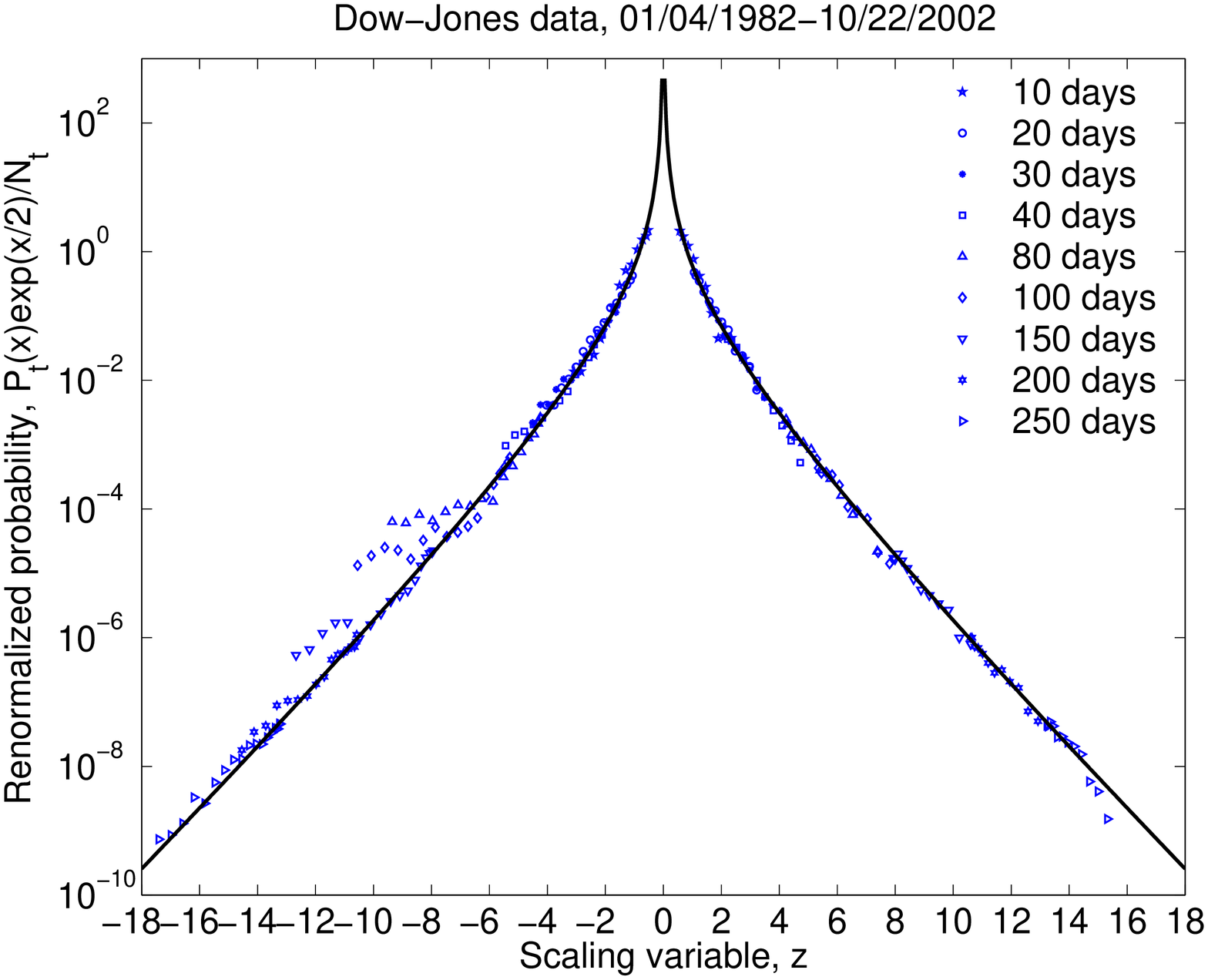,width=0.41\linewidth,angle=-90}}
\caption{\footnotesize\sf The same as in Fig.\ \ref{fig:DJ00} for
  1982--2002.}
\label{fig:DJ02}
\end{figure}

Similarly to Nasdaq, the S\&P500 data for 1982--1999 agree well with
the theory, as shown in Fig.\ \ref{fig:SP00}.  However, when the data
up to 2002 are added (Fig.\ \ref{fig:SP02}), deviations occur, albeit
not as strong as for Nasdaq.  For Dow-Jones 1982--1999 (Fig.\
\ref{fig:DJ00}), the data agrees very well with the theory.  The PDFs
for 1982--2002, shown in the left panel of Fig.\ \ref{fig:DJ02}, still
agree with the theory, but deviations are visible in the scaling plot
in the right panel of Fig.\ \ref{fig:DJ02}.  They come from the time
lags between 40 and 150 days not shown in the left panel.

\section{Discussion and conclusions}
\label{sec:conclusions}

We conclude that, overall, the PDFs of log-returns, $P_t(x)$, agree
very well with the DY formula \cite{Yakovenko} for all three
stock-market indices for 1982--1999.  It is important to recognize
that the single DY formula (\ref{Pfinal'-}) and (\ref{phaseF'-}) fits
the whole family of empirical PDFs for time lags $t$ from one day to
one year (equal to 252.5 trading days).  The agreement with the
nontrivial Bessel scaling function (\ref{Pbess-}) extends over the
astonishing {\it ten} orders of magnitude.  These facts strongly
support the notion that fluctuations of stock market are indeed
described by the Heston stochastic process.

On the other hand, once the data for 2000s are included, deviations
appear.  They are the strongest for Nasdaq, intermediate for S\&P500,
and the smallest for Dow-Jones.  The origin of the deviations can be
recognized by looking in Fig.\ \ref{fig:timeseries}.  Starting from
2000, Nasdaq has a very strong downward trend, yet we are trying to
fit the data using a constant positive growth rate $\mu$.  Obviously,
that would cause disagreement.  For S\&P500 and Dow-Jones, the
declines in 2000s are intermediate and small, so are the deviations
from the DY formula.  We think these deviations are not an argument
against the Heston model.  They rather indicate the change of $\mu$
from a positive to a negative value around 2000.  Our conclusion about
the change of regime is based on the statistical properties of the
data for the last 20 years.  The situation is very different from the
crash of 1987.  As our plots show, the crash of 1987 did not have
significant statistical impact on the PDFs of log-returns for 1980s
and 1990s, because the market quickly recovered and resumed overall
growth.  Thus, the crash of 1987 was just a fluctuation, not a change
of regime.  To the contrary, the decline of 2000s (which is
characterized by a gradual downward slide, not a dramatic crash on any
particular day) represents a fundamental change of regime, because the
statistical probability distributions have changed.  These conclusions
are potentially important for investment decisions.

The average growth rate $\mu$ is an exogenous parameter in the Heston
model and is taken to be constant only for simplicity.  In a more
sophisticated model, it could be a smooth function of time $\tau$,
reflecting the long-term trend of the market of the scale of 15--20
years.  Using a properly selected function $\mu_\tau$, one could
attempt to analyze the stock-market fluctuations on the scale of a
century.  That would be the subject of a future work.

We are grateful to Adrian Dr\u{a}gulescu for help and sharing his
computer codes for data processing and plotting.


\end{document}